# Gait Kinematics in Healthy Participants: A Motion Capture Dataset Under Weight Load and Knee Brace Conditions

Hanieh Moradi[1], Yas Vaseghi[1], Arash Abbasi Larki[1], Akram Shojaei[1], and Mehdi Delrobaei

*Faculty of Electrical Engineering, K. N. Toosi University of Technology, Tehran, 1631714191, Iran*
*Corresponding author: Mehdi Delrobaei (ORCID: 0000-0002-4188-6958), delrobaei@kntu.ac.ir*

A B S T R A C T

The objective assessment of gait kinematics is crucial in evaluating human movement, informing clinical decisions, and advancing rehabilitation and assistive technologies. Assessing gait symmetry, in particular, holds significant importance in clinical rehabilitation, as it reflects the intricate coordination between nerves and muscles during human walking. In this research, a dataset has been compiled to improve the understanding of gait kinematics. The dataset encompasses motion capture data of the walking patterns of eleven healthy participants who were tasked with completing various activities on a circular path. These activities included normal walking, walking with a weighted dominant hand, walking with a braced dominant leg, and walking with both weight and brace. The walking tasks involving weight and brace were designed to emulate the asymmetry associated with common health conditions, shedding light on irregularities in individuals' walking patterns and reflecting the coordination between nerves and muscles. All tasks were performed at regular and fast speeds, offering valuable insights into upper and lower body kinematics. The dataset comprises raw sensor data, providing information on joint dynamics, angular velocities, and orientation changes during walking, as well as analyzed data, including processed data, Euler angles, and joint kinematics spanning various body segments. This dataset will serve as a valuable resource for researchers, clinicians, and engineers, facilitating the analysis of gait patterns and extracting relevant indices on mobility and balance.

## 1. Introduction

The objective assessment of gait kinematics is essential for evaluating human movement, guiding clinical decisions, and advancing rehabilitation and assistive technologies (Tabor et al., 2021; Patterson et al., 2010, 2008). Assessing gait symmetry is particularly crucial in clinical therapy and rehabilitation engineering, as it reflects the complex coordination between nerves and muscles during human walking. Analyzing gait asymmetry can primarily be used to evaluate functional recovery and disease progression (Patterson et al., 2010; Böhm and Döderlein, 2012; Lundh et al., 2014; Gouwanda and Senanayake, 2011; Alves et al., 2020).

Though seemingly minor, the analysis of bilateral gait asymmetry yields critical insights into an individual's health status. Continuous monitoring of these asymmetries is vital for tracking patient progress or deterioration, providing valuable data for healthcare providers (Cabral et al., 2016). Additionally, evaluating subtle gait irregularities is essential for comparing the efficacy of treatments for neuromuscular disorders, thereby informing and optimizing therapeutic strategies (Kramers-de Quervain et al., 2004; Gouwanda and Senanayake, 2011; Anwary et al., 2018).

Collecting kinematic data on gait patterns under conditions that induce walking asymmetries can assist researchers in gaining a better understanding of the underlying mechanisms. Researchers and clinicians can utilize spatiotemporal variables to measure differences between the right and left sides of the body during walking tasks. These asymmetry measures provide insight into the quality of the walking pattern and offer essential information for gait analysis (Mills et al., 2013; Lewek et al., 2004; Patterson et al., 2010; Gouelle et al., 2018; Lewek et al., 2014).

For instance, the body joints' range of motion (ROM) is a fundamental yet effective parameter in analyzing body asymmetry. Joint ROM refers to the extent of movement at a joint necessary to displace a bone within its spatial limits. Restricted ROM in the upper and lower limbs can result in asymmetrical movement patterns between the left and right sides of the body, highlighting disparities in limb functionality. The clinical significance of ROM is profound, as it serves as an indicator of joint health and mobility (Clarkson, 2013; Shorter et al., 2008; Abu El Kasem et al., 2020; Hallaceli et al., 2014; Viteckova et al., 2018).

---

[1]These authors contributed equally to this work.

Consequently, healthcare professionals increasingly prioritize acquiring precise joint angle data to facilitate accurate assessments and comparisons. This detailed ROM analysis identifies specific impairments and informs targeted rehabilitation strategies, ultimately enhancing patient outcomes in clinical settings.

In a recent study, a dataset consisting of 138 healthy individuals and 50 stroke survivors was compiled by researchers using the Vicon motion capture system, ground-embedded force plates, and a synchronized surface EMG system to gather full-body kinematics (PiG-model), kinetics, and muscle activity of 14 back and lower limbs muscles. The study focused on capturing the participants' typical walking patterns during steady-state walking at self-selected speed without using walking aids or orthoses (Van Criekinge et al., 2023).

In another research, ten healthy individuals were recruited to participate in an experiment utilizing an inertial measurement unit (IMU) system and force plate in 33 combinations of average walking speed, step length, step frequency, and step width. The participants were required to walk on a split-belt instrumented treadmill (van der Zee et al., 2022).

The primary objective of the work presented by (Arippa et al., 2022) was to analyze the gait characteristics of 61 individuals with PD compared to 47 unaffected individuals through computerized 3D gait analysis using IR cameras and a treadmill. The findings indicated that patients with PD exhibited a modified gait pattern, particularly during the terminal stance and early swing phases of the gait cycle. Significant alterations in interlimb coordination at the ankle and hip joints were also observed.

In another study, amateur runners were enlisted to run on a running treadmill to investigate the effects of foot orthoses (FOs) on foot kinematics in the stance phase during running, equipped with 2 IMU modules attached to each foot. Results revealed that FOs reduced movement amplitude in all axes except for dorsi-plantar flexion in the left foot and both feet combined. Contact time and total step time increased significantly with FOs, while the number of steps remained unchanged (Florenciano Restoy et al., 2021).

Researchers have utilized IMU-based devices to assess gait symmetry during weight-bearing. In a research setting, eight healthy subjects were asked to hold weight in their hands while walking along a 12-meter lab walkway. The evaluation results suggested that most subjects exhibited constant or improved symmetry on the thighs relative to the shanks, with potential applications in wearable assistive device control or biofeedback gait retraining (Huan et al., 2020).

Upon reviewing the literature, we realized that the existing gait datasets have noticeable shortcomings, including a limited number of gait cycle samples, insufficient sensor coverage, and the lack of inclusion of raw data. Such deficiencies could make them unsuitable for comprehensive quantitative gait asymmetry monitoring and assessment. The proposed motion capture data is publicly available to address some of these limitations. It is also noteworthy that motion capture using IMU modules offers higher accuracy and reliability compared to video-based methods. The sensor modules employed in our data collection process provide ease of use due to their online data transmission and wireless functionality. This dataset aims to accelerate scientific progress in gait asymmetry assessment and monitoring by facilitating new and detailed analyses.

Our study involved eleven healthy participants who performed eight walking tasks on a circular path under various walking conditions, including normal walking, walking while carrying a weight in the dominant hand, walking with the support of a knee brace, and walking with both the weight and the brace. The purpose of selecting the tasks involving walking with a weighted dominant hand and a braced dominant leg was to simulate conditions relevant to gait asymmetry (Shorter et al., 2008; Abid et al., 2017). All the tasks were performed at both regular and accelerated speeds. Data was collected using ten wireless IMU modules, placed on relevant body parts to monitor joint angles and their derivatives with high temporal resolution. The prepared dataset includes both raw sensor outputs and processed information.

It is noted that, despite the relatively small sample size, the experimental design was crafted to capture a substantial number of gait cycles. The participants were all healthy individuals, which minimizes variability and ensures the reliability of the collected data. By structuring the tasks to elicit multiple repetitions of walking cycles, we aimed to enhance the robustness of the kinematic data. Given the homogeneity of the participant group, increasing the number of participants would likely not have provided significantly more insight.

Fig. 1 presents a process diagram that delineates the sequential steps undertaken in this study. The diagram provides a visual overview of the methodology, from participant recruitment and data collection to the validation phase.

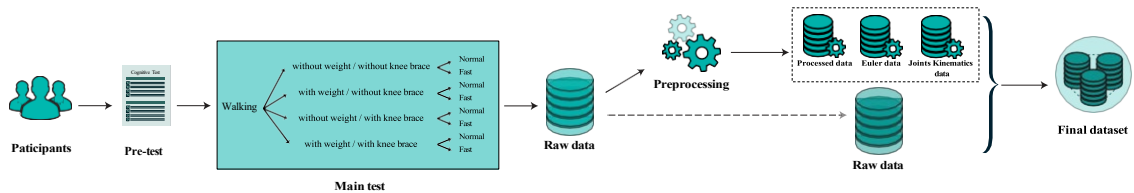

**Figure 1:** The diagram illustrating the data collection methodology employed in the proposed work.

**Table 1**
Demographic information of the participants.

| Demographic Variable | Value |
| --- | --- |
| Number of participants | 11, 5 female |
| Age (years) | 21.5 (3.0) |
| Height (m) | 1.71 (0.10) |
| Weight (kg) | 64.33 (9.24) |
| MMSE score | 29.13 |

## 2. Method

### 2.1. Participants

The dataset includes data from eleven healthy participants whose demographic information is demonstrated in Table. 1 The values are expressed as the mean throughout the manuscript, with the standard deviation in parentheses. All participants were asked to complete the Mini-Mental State Examination (MMSE) to ensure cognitive baseline consistency. Before participating in the study, all the participants signed a consent form agreeing to participate and authorizing the public release of the collected data. The consent form also included the following key points:

1. Participants received and understood all necessary explanations and instructions regarding the study and had the opportunity to ask any questions.
2. Participation in the study was voluntary, and participants understood they could withdraw from the research at any time without providing a reason.

### 2.2. Tasks

The participants were instructed to complete four laps of each of the following tasks along a predefined circular path, first at a regular pace and then at an accelerated speed:

1. *Normal Walk*
2. *Weighted Dominant Hand:* Walking while carrying a 1-kilogram weight in the dominant hand.
3. *Braced Dominant Leg:* Walking with a knee brace placed on the dominant leg.
4. *Combined Weight and Brace:* Walking with the 1-kilogram weight in the dominant hand and the brace on the dominant leg.

The designed tasks were intended to create an unequal distribution of mechanical effort on the arm and leg of the participants while walking in order to influence walking patterns. The tasks range from simple walking exercises to more complicated movements that challenge the balance and coordination of the participants. The data gathered from such tasks potentially gives insights into how differences in walking patterns can impact a person's ability to move and stay balanced.

Fig. 2 demonstrates a participant wearing the sensors and the knee brace while holding the weight.

### 2.3. Data Collection

The APEX motion capture system (Bonyan Sanat Novin Inc., Guilan, Iran) was employed for data collection. This system utilizes IMU modules, each equipped with gyroscopes, accelerometers, and magnetometers. Each IMU module in the system includes a battery and operates wirelessly. The IMU modules were securely attached using straps to various key points on the participants' bodies for accurate data capture, ensuring consistent positioning throughout the study.

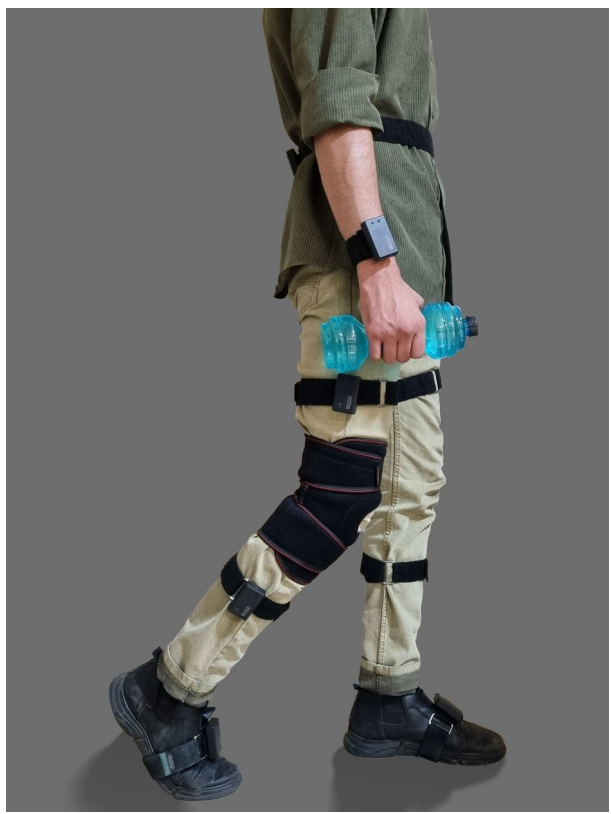

**Figure 2:** A participant wearing motion capture sensors, the knee brace, while holding the weight.

In this study, the participants wore a total of 10 IMU modules to monitor the corresponding joint angles during walking. Specifically, one sensor was attached to the trunk, one on the pelvis, one on each forearm, one on each thigh, one on each shin, and one on each foot.

The IMU modules were initially calibrated to ensure accurate data collection. The calibration procedure involved adjusting each module to ensure accurate measurement. Following calibration, the modules were synchronized to ensure that data transmission and reception were uniform across all modules, allowing them to send data simultaneously. The modules communicate with the software using a modem, connecting to a laptop wirelessly or via a LAN cable.

Once the modules were calibrated and synchronized, they were installed on the participants' bodies. Participants were asked to stand still and upright to allow the software to adjust the modules relative to the reference. This step ensured the accurate alignment of the sensors.

These sensors are designed to measure joint angles relative to a specified reference point during walking. For instance, the reference point for the legs is the pelvis, and the reference point for the upper body is the trunk. The sensors function with a sampling rate of 4 milliseconds, significantly enhancing the resolution of the measurements. The system processes, refines, and transmits the data automatically, ensuring clean and accurate data collection.

All data was collected from 3 PM to 7 PM to ensure consistency among participants. In case a participant felt fatigued between tasks, they were allowed to rest for one to two minutes to prevent data distortion due to fatigue.

The dimensions of the path along which the tasks were performed are shown in Fig. 3 .

The modules were installed on various parts of the body as follows:

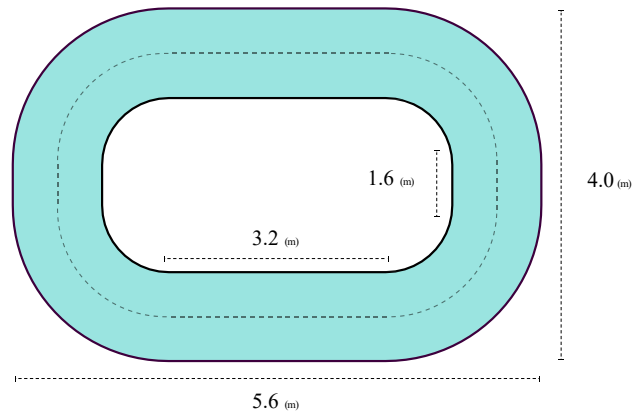

**Figure 3:** The circular path along which the participants performed the tasks.

- *Trunk and pelvis:* The modules on these parts were installed in the posterior view, on the frontal plane, and aligned to ensure the Y-axis was perpendicular to the ground.

- *Forearms, hands, thighs, and shins:* The modules were installed in the lateral view, on the sagittal plane, and aligned to ensure the Y-axis was perpendicular to the ground.

- *Feet:* The modules were installed on the metatarsal bones, aligned with the longitudinal axis of the limbs.

In Fig. 4 the placement of the modules on each participant's body is shown.

## 2.4. Data Validation

The validity and reliability of the data collection process have been thoroughly evaluated to ensure its applicability.

1. *Calibration and Synchronization:* Accurate data collection using IMU depends significantly on proper calibration and synchronization of the sensors. We calibrated each IMU module to ensure precise measurement of accelerations, angular velocities, and magnetic fields. The calibration process included adjusting for sensor biases and alignments to mitigate errors during data capture. Sensor synchronization was performed to ensure temporal alignment across all modules, preventing time-skewed data.
2. *Data Quality and Integrity:* To assess the integrity of the collected data, we implemented multiple data validation techniques. First, we conducted a static position test to confirm that the sensors were correctly zeroed. During the test, participants stood still, and sensor readings were observed for any drifts or fluctuations that might indicate calibration issues. Subsequently, dynamic validation tests, including repeated movement patterns, were performed to check for consistency in sensor readings.
3. *Gait Task Design:* The structured tasks designed for this study, involving different walking speeds and conditions, were implemented to simulate various real-world scenarios. These tasks were informed by previous studies that have utilized similar protocols to examine gait asymmetry and balance (Shorter et al., 2008; Dong et al., 2022). This comprehensive approach ensures that our dataset captures a wide range of gait characteristics, making it suitable for diverse research applications.

## 3. Data access

All datasets generated during the current study are publicly available on figshare (Shojaei et al., 2024) in CSV files organized by subjects and trials. Each trial includes 4 CSV files containing raw sensor outputs and analyzed information from the IMU modules as follows:

1. *Raw Sensor Data:* The raw data file contains the raw outputs from sensor installed on different body parts. Each row represents data from one sensor at a given time point. The first row indicates the sensor names corresponding to body parts.
   For each sensor, the data is organized as follows:

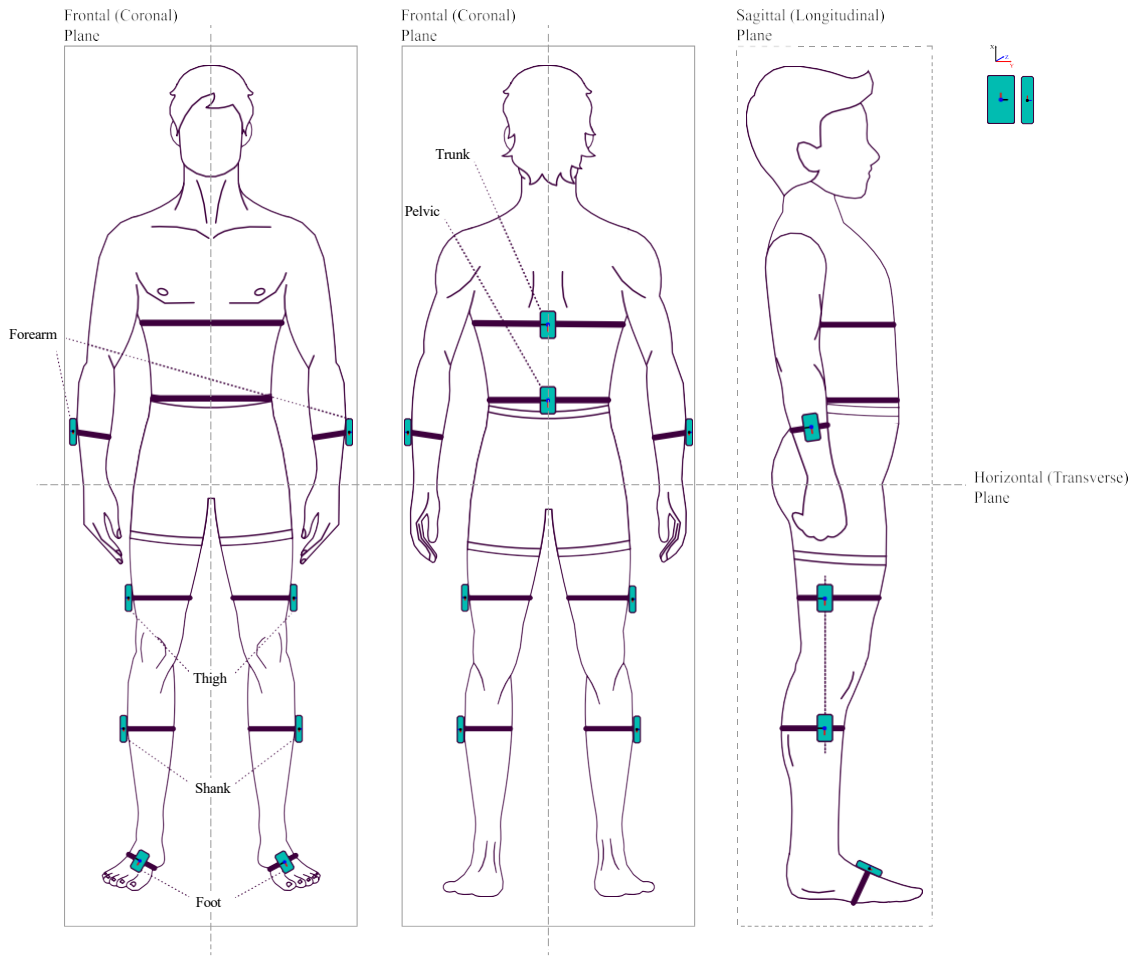

**Figure 4:** The placement of the IMU modules on relevant body parts.

- *Time Data:* The time of recording each point in milliseconds.
- *Acceleration Data:* Contains the raw data from the accelerometer sensor on the X, Y, and Z axes of the IMU modules.
- *Gyroscope Data:* Includes the raw data from the gyroscope sensor on the X, Y, and Z axes of the IMU modules.
- *Magnetometer Data:* Contains the raw data from the magnetometer sensor on the X, Y, and Z axes of the IMU modules.

2. Analyzed Data:
    - *Processed Data:* Contains quaternion components, acceleration in the IMU coordinate system, linear acceleration without gravity, acceleration in the global coordinate system, and time.
    - *Euler Angles Data:* Includes roll, pitch, and yaw angles which describe the orientation of the body segments.
    - *Joints Kinematics Data:* Contains relative angles of joints such as abduction-adduction, internal-external rotation, and flexion-extension.
        - *Abduction-Adduction:* Related to the changes in the joint angles in the frontal plane.
        - *Internal-External Rotation:* Related to the changes in the joint angles in the horizontal plane.
        - *Flexion-Extension:* Related to the changes in the joint angles in the sagittal plane.

Table. 2 to Table. 5 present the description of the collected data and Table. 6 demonstrates the various body parts utilized in the collection of each set of data.

## 4. Discussion

In this study, we collected data from healthy participants under different walking conditions to construct a comprehensive dataset for future analysis of gait asymmetry. Earlier studies have evaluated gait asymmetry using various kinematic data but often did not publish their datasets, limiting access to further research. These studies also had limitations, such as a limited number of gait cycle samples, insufficient sensor coverage, lack of raw data, and low sampling rates, especially in video-based methods (Van Criekinge et al., 2023; van der Zee et al., 2022; Arippa et al., 2022; Florenciano Restoy et al., 2021; Huan et al., 2020; Scherpereel et al., 2023).

Our dataset includes all necessary data for analyzing gait asymmetry, such as raw sensor outputs, Euler angles, and kinematic data. Video-based methods often suffer from low accuracy compared to IMU-based methods. The fact that all our participants were healthy contributes to the high accuracy of the collected data. Moreover, this dataset contains a sufficient number of gait cycles to enhance reliable results on asymmetry analysis. The availability of raw and processed data facilitates a wide range of analyses, making the dataset a versatile resource for advancing the field of gait analysis.

To our knowledge, no published dataset combines the specific walking tasks and conditions we employed in healthy participants, making our dataset unique and valuable for future research. Most existing datasets focus solely on lower or upper limb data, whereas ours includes both, providing a more comprehensive analysis of gait asymmetry.

A key strength of this dataset is its detailed capture of both upper and lower body kinematics at different walking speeds. This allows for a thorough examination of how gait patterns change under various conditions and speeds, providing a richer understanding of the dynamics involved. Wireless IMU modules enhance the data's accuracy and reliability, making it a robust tool for researchers and clinicians.

The findings can potentially impact clinical practices. The detailed joint dynamics and angular velocities captured can help identify specific impairments and tailor rehabilitation programs to address these issues. The ability to monitor changes in gait symmetry over time also provides a valuable metric for assessing the effectiveness of therapeutic interventions.

This dataset addresses several limitations compared to previous studies, such as the inclusion of a larger number of gait cycles and comprehensive sensor coverage. The high temporal resolution of the data further enhances its utility, allowing for precise tracking of joint movements and their coordination. This level of detail is essential for understanding the subtle nuances of gait asymmetry and developing targeted interventions.

Despite our dataset's strengths, this study has some limitations. IMU-based motion capturing may have restrictions in capturing complex motions and environmental factors, which can influence gait parameters. Future studies should consider incorporating more diverse participants and exploring additional walking conditions to enhance the dataset's robustness.

The implications of this research extend beyond clinical rehabilitation. The insights gained from analyzing gait kinematics can inform the design of assistive devices and wearable technologies. Understanding how different conditions affect gait can lead to the development of more effective prosthetics and orthotics that better mimic natural movement patterns.

## 5. Conclusion

This work presented a motion capture dataset focusing on gait kinematics in healthy participants under varying weight load and knee brace conditions, offering insights into human locomotion. The dataset includes raw sensor data and processed data, facilitating detailed studies of gait patterns and asymmetry. Its accessibility and the inclusion of diverse walking conditions make it a crucial resource for researchers, clinicians, and engineers involved in gait analysis, rehabilitation, and the development of assistive technologies. By addressing the limitations of existing gait datasets, the proposed dataset aims to accelerate progress in gait kinematics assessment and monitoring. The detailed analyses and insights it provides can enhance understanding of gait symmetry and the effects of different loading conditions on human walking. Furthermore, it has the potential to inform clinical evaluations, rehabilitation methods, and the development of assistive technologies, ultimately contributing to improved patient outcomes related to mobility, balance, and rehabilitation.

**Table 2**
Raw Data Parameters and Descriptions.

| Parameter | Description |
| --- | --- |
| Time | Timestamp for each recorded data point |
| AccX | Accelerometer data along X-axis |
| AccY | Accelerometer data along Y-axis |
| AccZ | Accelerometer data along Z-axis |
| GyroX | Gyroscope data along X-axis |
| GyroY | Gyroscope data along Y-axis |
| GyroZ | Gyroscope data along Z-axis |
| MagX | Magnetometer data along X-axis |
| MagY | Magnetometer data along Y-axis |
| MagZ | Magnetometer data along Z-axis |

**Table 3**
Processed Data Parameters and Descriptions.

| Parameter | Description |
| --- | --- |
| Time | Timestamp for each recorded data point |
| Q0 | Quaternion Scalar Component |
| Q1 | Quaternion Vector Component |
| Q2 | Quaternion Vector Component |
| Q3 | Quaternion Vector Component |
| Acc_X | Acceleration in IMU coordinate system along X-axis |
| Acc_Y | Acceleration in IMU coordinate system along Y-axis |
| Acc_Z | Acceleration in IMU coordinate system along Z-axis |
| Acc_linX | Linear acceleration without gravity along X-axis |
| Acc_linY | Linear acceleration without gravity along Y-axis |
| Acc_linZ | Linear acceleration without gravity along Z-axis |
| Acc_GlinX | Acceleration in the global coordinate system along X-axis |
| Acc_GlinY | Acceleration in the global coordinate system along Y-axis |
| Acc_GlinZ | Acceleration in the global coordinate system along Z-axis |

**Table 4**
Euler Angles Parameters and Descriptions.

| Parameter | Description |
| --- | --- |
| Time | Timestamp for each recorded data point |
| Roll | Roll angle |
| Pitch | Pitch angle |
| Yaw | Yaw angle |

# Reference Data

All datasets produced in this study are accessible to the public on Figshare (Shojaei et al., 2024).

# Declaration of Competing Interest

All authors declare that they have no conflicts of interest.

# Acknowledgment

This work was supported by K. N. Toosi University of Technology's Chancellor's Visionary Grant.

**Table 5**
Joints Kinematics Parameters and Descriptions.

| Parameter | Description |
|---|---|
| Time | Timestamp for each recorded data point |
| Abduction-Adduction | Abduction-Adduction angle |
| Internal-External Rotat | Internal-External Rotation angle |
| Flexion-Extension | Flexion-Extension angle |

**Table 6**
Body Parts for Different Data.

| Body Parts for Raw, Processed, and Euler Data | Body Parts for Joints Kinematics Data |
|---|---|
| LeftFoot | LeftAnkle |
| RightFoot | RightAnkle |
| LeftShank | LeftKnee |
| RightShank | RightKnee |
| LeftThigh | LeftHip |
| RightThigh | RightHip |
| LeftHumerus | LeftShoulder |
| RightHumerus | RightShoulder |
| Pelvic | Pelvic |
| Trunk | Trunk2Ground |